\documentclass{article}[11pt]

\usepackage[a4paper, total={6in, 8.5in}]{geometry}
\usepackage{hyperref}

\usepackage{graphicx}%
\usepackage{multirow}%
\usepackage{amsmath,amssymb,amsfonts}%
\usepackage{amsthm}%
\usepackage{mathrsfs}%
\usepackage[title]{appendix}%
\usepackage{xcolor}%
\usepackage{textcomp}%
\usepackage{manyfoot}%
\usepackage{booktabs}%
\usepackage{algorithm}%
\usepackage{algorithmicx}%
\usepackage{algpseudocode}%
\usepackage{listings}%

\newcommand{\nn}{\nonumber}

\def\be{\begin{equation}}
\def\ee{\end{equation}}

\def\bea{\begin{eqnarray}}
\def\eea{\end{eqnarray}}

\def\ba{\begin{aligned}}
\def\ea{\end{aligned}}
\def\bt{\begin{tabular}}        
\def\et{\end{tabular}}
\def\bc{\begin{center}}        
\def\ec{\end{center}}
\def\bi{\begin{itemize}}
\def\ei{\end{itemize}} 

\begin{document}

\title{\bf Scale hierarchies near the conifold}

\author{Nana Cabo Bizet\thanks{nana@fisica.ugto.mx} \and Oscar Loaiza Brito\thanks{oloaiza@fisica.ugto.mx} \and Yessenia Olguín Trejo\thanks{yess27.olt@gmail.com}}

\date{
    \begin{minipage}{\textwidth} \it 
        \centering
           Departamento de F\'isica, Universidad de Guanajuato, Loma del Bosque 103,  \\ Le\'on, 37150, Guanajuato, M\'exico
    \end{minipage}
    \vspace{1cm}
}

\maketitle

\begin{abstract}
We study the axio-dilaton and complex structure stabilization for a generic one-parameter Calabi-Yau (CY) compactification. We focus on near the conifold regions and consider a strongly warped metric. We analyze numerically the case of the mirror of the quintic CY. The axio-dilaton $\tau$ and complex structure $z$ moduli are stabilized simultaneously comparing with previous results, showing that generically $z$ is heavier than $\tau$, and the axio-dilaton can not be fixed first. This is also the case when we add the warping correction to the potential; and in general this inclusion does not destabilize the moduli. We point out why in this setup non-supersymmetric vacua are much more dense than supersymmetric vacua. We study the dependence of the vacua stabilization by addition of an $\overline{D3}$-brane, for a fixed volume. The vacuum state masses have divergent eigenvalues near the conifold singularity that gets softened by the warping correction. Generically the considered contributions allow to find stable vacua in this setup. The scale hierarchy between 6D and 4D; arising from the warping, as stated by Giddings Kachru and Polchinski (GKP), is preserved; while the masses in the vacua acquire physically reasonable values.
\end{abstract}

\section{Introduction}
String theory vacua have been explored actively in the last decades. Warped compactifications constitute an interesting region in the string landscape. They provide a natural hierarchy between the four dimensional (4D) scale and the internal six dimensional (6D)  scale. This fact was  noticed in the relevant work by Giddings, Kachru and Polchinski (GKP) \cite{Giddings:2001yu}, and they were until recent years compelling candidates to populate the {\it de Sitter} portion of the landscape. When the effects of global Calabi-Yau geometry are taken into account in the effective theory, the scale hierarchies can get modified \cite{CLZ}.  In recent years, the picture established in GKP has been challenged for many reasons (\cite{Danielsson:2018ztv,Sethi:2017phn,Bena:2020xrh} and references thereof). One of this, lays foundation on the fact that the warping can give rise to important modifications to the stabilization potential of the moduli \cite{Douglas:2007tu,Douglas:2008jx,Grana, Blumenhagen:2019qcg}\footnote{ The work of \cite{Lust:2022xoq} states that there could be new modifications to the scalar potential due to the Einstein's equations.}. In the literature where warped compactifications have been considered, employing the GKP solution for the complex structure, the no-scale structure of the potential is preserved by the warping correction to the K\"ahler potential. Several works have followed the observations in \cite{Grana}, exploring the effects of the warping on the moduli stabilization mechanism of type IIB fluxes setup \cite{Buratti:2018onj,Bena:2019sxm,Dudas:2019pls,Seo:2021kyi,Seo:2023pyp,ValeixoBento:2023nbv,Seo:2023ssl,Lust:2022mhk,Tsagkaris:2022apo,Blumenhagen:2022dbo,Gao:2022uop,ValeixoBento:2022qca,Bento:2021nbb,Blanco-Pillado:2020hbw,Gao:2020xqh,Scalisi:2020jal,Parameswaran:2020ukp}. As well the exploration of the moduli stabilization mechanism close to the conifold singularity without considering the warping corrections continues to be explored \cite{Alvarez-Garcia:2020pxd,Demirtas:2020ffz}. We will investigate here how this scale hierarchy is affected by the warping corrections and by the inclusion of an anti-D3 brane.

We consider to this aim the effect of the warping in the scalar potential arising from flux compactifications of type IIB string theory. The novelty is the specification of the compactness of the internal dimensions, and the input for the periods of a concrete model, which
for the numerics we take to be the mirror of the quintic in $\mathbb{P}^4$ \cite{CLZ}.
The work developed in \cite{CLZ} by two of the authors sets the basis for the exploration. The setup is 10D type IIB string theory with fluxes, and the scalar fields are: the axio-dilaton and the single complex structure (cs). Our analytical results apply moreover to CY with a single complex structure parameter, but for the numerical explorations we choose the mentioned geometry.
We compute the effective scalar potential, and obtain that the axio-dilaton has to be stabilized together 
with the cs modulus; this occurs because the axio-dilaton is generically lighter than the cs modulus. The cs modulus singular value 
gives rise to the conifold geometry; we find that there can be stable compact  cases where the scale hierarchies found by GKP \cite{Giddings:2001yu} are preserved. We incorporate the effect of the warping in the potential \cite{Grana} and analyze the consequences for the vacua. Finally we study the effect
of the uplifting of the scalar potential via an anti-D3 brane, finding positive results for this test.

The work is organized as follows: In Section \ref{sec2} we present the solutions for Minkowski vacua in type IIB compactifications on Calabi-Yau(CY) varieties with one cs modulus, without warping corrections. In Section \ref{sec3}  we discuss the modifications 
of the vacua by including the warping corrections in the scalar potential. In Section \ref{sec4} we present the
uplifting of the scalar potential given by anti D3-branes.  In Section \ref{sec5} we present our conclusions and ideas for further work. In the single Appendix \ref{appA} we develop formulae for the mass matrix for vacua near the conifold, that can be useful for following investigations.

\section{Vacua without warping correction}
\label{sec2}
In this Section we present the solutions for vacua near the conifold, when no warping correction is included into the K\"ahler potential.  The warped metric,  which describes the internal 6D, and that preserves the Poincar\'e symmetry reads \cite{Giddings:2001yu}

\begin{eqnarray}\label{metric}
ds^2_{10} = e^{2A(y)}\eta_{\mu\nu}dx_{\mu}dx_{\nu} + e^{-2A(y)}\tilde g_{mn}dy^mdy^n.
\end{eqnarray}
$\mu,\nu$ denote space-time indices and $a,b$ denote the internal indices. The hierarchy between the spacetime (4D) and compactification (6D) physical scales will be given by the distance of the vacuum  to the conifold $z_0$ (i.e. the vev of the cs modulus) as $e^{A}\sim z_0^{1/3}$ \cite{Giddings:2001yu,comments}.

The scalar potential in flux compactifications of type IIB string theory can be written in an ${\cal N} =1$ supergravity form as
\be
V =\frac{1}{2\kappa_{10}^2 g_s}\,e^K\left[K^{a\bar b}D_a W \overline{D}_{\bar b} \overline{W} -|W|^2 \right],\label{GVW}
\ee
$g_s$ is the string coupling and the constant $\kappa_{10}=\frac{l_s^8}{4\pi}$ depends on the string length. The scalar potential (\ref{GVW}) depends on the superpotential $W$ and the K\"ahler potential $K$.
The indices $a,b$ denote the moduli fields, $K^{a\bar b}$  is the inverse metric in field space and $D_a W=\partial_a W+\partial_a K W$ is the supersymmetric covariant derivative of $W$. The superpotential generated by the fluxes is the Gukov-Vafa-Witten (GVW) superpotential \cite{Gukov:1999ya}:  
\begin{eqnarray}
\label{fluxW}
W &=& \int_{CY} G_{(3)}\wedge \Omega=   \int_{CY}( F_{(3)}- \tau  H_{(3)})  \wedge \Omega
=G \,\Sigma\, \Pi  \,,
\end{eqnarray}
where $\Sigma$ denotes the symplectic matrix, and $\Pi$ denotes the periods of the CY, defined by the integrals of the holomorphic $(3,0)$ form of the CY,   $\Omega$, over a canonical homology basis. This homology basis parametrises the complex structure moduli $z^{\alpha}$, in the present case a single one $z$. For the case of one complex structure modulus, the dimension of homology is given by the Betti number $b_3= h_{2,1}+h_{1,2}+h_{3,0}+h_{0,3}=4$. We employ a symplectic basis for forms $(\alpha_I,\beta^I)$ in cohomology $H^3(CY,\mathbb{Z})$,and for cycles $(A^I,B_I)$ in homology $H_3(CY,\mathbb{Z})$. The forms and cycles satisfy:
\begin{eqnarray}
    \int_{CY}\alpha_I\wedge \beta^I&=&\delta_I^J, \\ \int_{CY}\alpha_I\wedge \alpha_J&=&\int_{CY}\beta^I\wedge \beta^J=0, \\
    \int_{A^J}\alpha_I&=&-\int_{B_I}\beta^J=\delta_I^J.
\end{eqnarray}

Thus the period vector has 4 components. The K\"ahler potential is given by 
\be\label{kahler}
K= -\ln \left[-i \left(\tau-\bar \tau \right)\right] -\ln \left[i \int_{CY} \Omega\wedge \bar \Omega\right] - 2\ln \left[{\cal V} \right]\,, 
\ee
where $\cal V$ is the dimensionless volume defined in terms of the dimensionless K\"ahler moduli $T_m$ and $\tau = C_0 +i \,e^{-\phi}=t_1+i\, t_2$ (where $e^{\phi} = g_s$ defines the string coupling). We will use the mirror of the quintic on $\mathbb{P}^4$ manifold to test some of the results in \cite{Grana}. Periods for this manifold have been calculated previously in \cite{Candelas:1990pi}, in \cite{CLZ} we obtained them for an arbitrary order, to study the convergence of the vacua solutions. We use the following notation for the  components  of the periods and the fluxes:
\begin{eqnarray}
\Pi=\binom{\mathcal{X}^I}{\mathcal{F}_I}  
&=&\left(
\begin{array}{c}
\Pi_1\\\Pi_2\\\Pi_3\\\Pi_4\end{array}\right),  \, \\
(F_{(3)}^I,F_{(3)I})&=&(F_1,F_2,F_3,F_4),\, \nn \\
(H_{(3)}^I,H_{(3)I})&=&(H_1,H_2,H_3,H_4), \nn\\
(G_{(3)}^I,G_{(3)I})&=&(G_1,G_2,G_3,G_4), \\
\Pi&=& \mathcal{X}^I \alpha_I -\mathcal{F}_I\beta^I, \nn \\
F_{(3)}&=& F_{(3)}^I \alpha_I -F_{(3)I}\beta^I, \nn \\
H_{(3)}&=& H_{(3)}^I \alpha_I -H_{(3)I}\beta^I, \nn
\end{eqnarray}\label{flujos}

where the fluxes satisfy the Dirac quantization, 

\begin{eqnarray}
 \frac{1}{(2\pi)^2\alpha'}\int_{A^I}F_{(3)}&=&M^I,\, \, \,   \frac{1}{(2\pi)^2\alpha'}\int_{A^I}F_{(3)}=M_I, \\ 
  \frac{1}{(2\pi)^2\alpha'}\int_{A^I}H_{(3)}&=&N^I,\, \, \,  \frac{1}{(2\pi)^2\alpha'}\int_{A^I}H_{(3)}=N_I,
\end{eqnarray}
with $M_I$, $M^I$, $N^I$, $N_I$ integers.\\

Let us comment here on the tadpole condition the fluxes must fulfill. Since we are considering a type IIB flux compactification, there are contributions from 3-dimensional branes and orientifolds as well as 7-dimensional ones. Specifically
\begin{eqnarray}
 N_{flux}&=&\frac{N_{O3}}{2}-2 N_{D3}+\frac{\chi(D_{O7})}{6}+\sum_a N_a \frac{\chi(D_a)}{12}, \label{tadpole3}\\
8[D_{O7}]&=&\sum_aN_a([D_a] + [D_a' ]),\label{tadpole7}
\end{eqnarray}
where
\begin{eqnarray}
N_{flux}&=&\frac{1}{\kappa_{10}^2T_3}\int_{CY}H_{(3)}\wedge F_{(3)}.\label{Nflux}
\end{eqnarray}

The D3 tadpole condition is (\ref{tadpole3}) and the D7 tadpole condition is (\ref{tadpole7}). $N_{flux}$ in (\ref{Nflux}) is the fluxes constribution to the D3 tadpole. $D_a$ denotes a divisor wrapped by a D7 brane,  $\chi(D_a)$ the Euler number of the divisor, and $N_a$ the number of $D7$ branes on $D_a$.
The divisor classes are $[D_a]$, $[D_a']$ and $[D_{O7}]$. We are interested in a configuration suggested in \cite{Denef:2004ze}. This is the D7 branes are on top of the O7-plane, and they satisfy $\sum_a N_a=4$ and the divisors are then $[D_a]=[D_a']=[D_{O7}]$. This guaranties that there is no contribution from the D7 to the scalar potential. In this family of solutions we have only the tadpole condition:
\begin{eqnarray}
 N_{flux}&=&\frac{N_{O3}}{2}-2 N_{D3}+\frac{\chi(D_{O7})}{2}. \label{tadpole3OK}
 \end{eqnarray}

Fulfilling this condition requires an interplay among the number of $D3$-branes, the flux number $N_{flux}$, and the Euler number $\chi(D_{O7})$. In the following, we shall concentrate only on the flux contribution to the scalar potential, assuming that we can always fulfill the tadpole contribution by adding as many $D3$-branes as necessary to balance the contribution from fluxes and the Euler number (which we have not computed). Once the tadpole is satisfied, the contribution of $D3$-branes to the scalar potential vanishes, and consequently, we shall not consider it.\\


 Hence, we shall consider a no-scale potential for the Kähler moduli
such that $K^{i\bar j}D_i W \overline{D}_{\bar j} \overline{W} -|W|^2=0$, with indices $i$ and $j$ running over the K\"ahler moduli. Assuming that the scalar potential is positive definite as:

\be
V =\frac{1}{2\kappa_{10}^2 g_s}\,e^K\left[K^{a\bar b}D_a W \overline{D}_{\bar b} \overline{W} \right],\label{Vnoscale}
\ee
where now $a,b$ run over the axio-dilaton and the complex structure modulus. The only dependence on the K\"ahler moduli $T_m$ lays on $K$.

\subsection{Minkowski vacua equations}

We describe next how the Minkowski vacua of the potential (\ref{Vnoscale}) i.e. $V=0$ can be obtained, this is based on \cite{CLZ}. We require the vanishing of the covariant derivative of the superpotential $W$ with respect to the axio-dilaton: $D_{\tau}W=0$. This leads to a dilaton profile $\tau_\tau(z)$; as well we require the vanishing of the covariant derivative $D_{z}W=0$ leading to a dilaton profile $\tau_z(z)$. Both solutions have to be equal  \cite{CLZ}:

\be\label{tT}
\tau_\tau(z) = \frac{F\Sigma\bar \Pi}{H\Sigma\bar \Pi},\, \\ \, \,  \,
\tau_z(z) = \frac{F\Sigma(\widetilde{\Pi} \Sigma\bar \Pi)}{H\Sigma(\widetilde{\Pi}\Sigma\bar \Pi)}, \, \,  \,  \tau_\tau(z)=\tau_z(z).
\ee
In previous formula $\widetilde \Pi=\partial_z \Pi  \otimes \Pi-\Pi  \otimes \partial_z \Pi$ ($\widetilde \Pi$ has two indices).  The complex structure is fixed $z=z_0$ and $\tau(z_0)=\tau_0$. This is generically a  Minkowski non-supersymmetric vacuum with $V=0$ provided  $W\neq 0$. Having a supersymmetric vacuum would imply in addition to have $W=0$. We follow the  strategy of \cite{CLZ} to find Minkowski vacua by setting the real and imaginary parts  of $\tau_\tau-\tau_z$  to zero. We take the numerator of this quantity,
avoiding fractional values of polynomials in $z$ and $\ln z$, this is written as
\begin{align}
A_{0}&=(H\Sigma\bar\Pi)(F\Sigma(\widetilde\Pi\Sigma \bar\Pi))-(F\Sigma\bar\Pi)(H\Sigma(\widetilde\Pi\Sigma \bar\Pi)).\label{ecu}
\end{align}

Let us emphasize something here, as mentioned these are in general non-supersymmetric vacua \cite{CLZ}. Supersymmetric vacua in addition would require $W=G\Sigma\Pi=(F\Sigma\Pi)-\tau (H\Sigma \Pi)=0$. This implies that one needs to solve for the same value of $z_0$ two complex equations (\ref{ecu}) and (\ref{fluxW}). Thus, we conclude that non-SUSY vacua are much more dense than SUSY vacua in this string theory setup. Notice that recently, those more scarce SUSY vacua haven been found in \cite{Candelas:2023yrg} employing the modularity characteristics of  Calabi-Yau 3-folds \cite{Kachru:2020sio}. To explore these vacua in the context of realistic type IIB string compactifications, one would need to include the warping corrections \cite{Douglas:2007tu,Douglas:2008jx,Grana}, we implement here for the compact case.

\subsection{GKP fluxes/solution} 
In this subsection we write the vacuum expectation values for the complex structure modulus ($z_0$) and the axio-dilaton ($\tau_0$), by stabilizing the moduli near to the conifold.  This text constitutes a review of the work performed in \cite{CLZ}.This is done for the case discussed in GKP \cite{Giddings:2001yu} where only three fluxes are on.

We write the solution for $z_0$ and $\tau_0$ when only the fluxes $F_1,H_3$ and $H_4$  are on and $H_3 \gg H_4$.   Let us define the notation: $H_3=K$, $F_1=M$ where $K,M$ are used in \cite{Giddings:2001yu,Grana}.  As discussed in \cite{Giddings:2001yu} one  needs a flux, $H_4$, in order to stabilise the dilaton (note however that in \cite{Grana} their final potential only includes $K,M$, as they assumed other fluxes stabilising $\tau$ are encoded in $g_s$).  The  superpotential  for this case is given by 
\begin{eqnarray}
W=F_1 \Pi_3+\tau(H_3\Pi_1+H_4\Pi_2).
\end{eqnarray}
The  value of the axio-dilaton arises from $D_{\tau} W=\partial_z W+K_z W=0$,
with $K_z=\alpha \frac{}{}$ is given by
\begin{eqnarray}
\tau_0=-\frac{F_1\bar \Pi_3^0}{H_4 \bar \Pi_2^0}.\label{tGPK}
\end{eqnarray}
The parameters are  $\Pi_i^0=\lim_{z\to 0}\Pi_i$ and $\partial_z{\Pi}_i^0=\lim_{z\to 0}\partial_z{\Pi}_i$ , i.e. the values of the periods at  $z=0$ (the conifold point). The third period component near the conifold reads $\Pi_3=\Pi^0_3+\alpha z+\beta z \ln z+O(z)$. 
To estimate  the complex structure value at the minimum, $z_0$, we consider the leading terms in $D_z W=0$. We compute $\partial_z W$ evaluated at (\ref{tGPK})  and $D_z W$ to obtain:
\begin{eqnarray}
\partial_z W&=&F_1(\alpha+\beta+\beta\ln z)-\frac{F_1\bar \Pi_3^0}{H_4 \bar \Pi_2^0}(H_3 \partial_z{\Pi}_1^0+H_4\partial_z{\Pi}_2^0)+O(z).\nn \\
D_z W&=& F_1\beta\ln z+\tau_0H_3 \partial_z{\Pi}_1^0+a_0+O(z),\\
W_0&=&(F_1 \Pi_3^0+\tau_0(H_3\Pi^0_1+H_4 \Pi^0_2)),\nn \\
\partial_z K_0&=&- \epsilon(\bar\Pi^0_2\partial_z{\Pi}^0_4-\bar\Pi^0_4\partial_z{\Pi}^0_2-\bar \Pi_3^0\partial_z{\Pi}^0_1)/(\bar\Pi^0_2\Pi^0_4-\bar\Pi^0_4\Pi^0_2)
,\nn \\
a_0&=&F_1(\alpha+\beta)+\tau_0 H_4\partial_z{\Pi}^0_2+\partial_z K_0 W_0. \nn
\eea
A parameter $\delta_0$ is defined \cite{CLZ}, which allows to measure the departure from the formulas in \cite{Giddings:2001yu}:
\bea
\delta_{0}&=&\frac{a_0}{F_1}=\alpha+\beta-\frac{\bar\Pi^0_3}{\bar\Pi^0_2}\partial_z{\Pi}^0_2+\partial_z K_0 \Pi^0_3-\partial_z K_0\frac{\bar\Pi^3_3}{\bar \Pi^0_2}\Pi^0_2\,, \nn
\end{eqnarray}
where we used \eqref{tGPK} and $W_0$ in $a_0$.  Substituting $\Pi_1$ at the conifold point, 
 $\Pi^0_1=0$, neglecting  $O(z)$  and  $\delta_0$  terms (which are $O\left(\frac{H_3}{H_4}\right)$) one gets\footnote{The derivative of $K$ reads $\partial_z K=-\frac{\bar \Pi^T\Sigma \partial_z \Pi}{\bar \Pi^T \Sigma \Pi}$, and closed to the conifold its more relevant contribution would come from $\bar \Pi_1 \partial_z\Pi_3\sim \bar \Pi^0_1 \beta \ln z $, but $\bar \Pi_1^0=0$. 
Therefore the  most relevant contribution is the constant term $\partial_z K_0$.} 
\begin{eqnarray}
z_{old}=\exp \left(\frac{H_3}{H_4}\frac{\partial_z{\Pi}^0_1\bar\Pi^0_3}{\beta\bar{\Pi}^0_2}\right)=\exp \left(-\tau_0\frac{H_3}{F_1}\frac{\partial_z{\Pi}^0_1}{\beta}\right).\label{zH}
\end{eqnarray}
This is the result of  GKP \cite{Giddings:2001yu}.  
Including the correction of $\delta_0$,  \eqref{zH} the cs at the Minkowski minimum becomes 
 \begin{eqnarray}
z_{new}=\exp \left(\frac{H_3}{H_4}\frac{\partial_z{\Pi}^0_1\bar\Pi^0_3}{\beta\bar{\Pi}^0_2}-\frac{\delta_{0,\epsilon}}{\beta}\right).\label{zHok}
\end{eqnarray}
The extra factor   $\exp\left(-\frac{\delta_0}{\beta}\right)\sim 20$ with respect to GKP, appears after considering originally neglected terms contributing to $D_z W=0$. These corrections due to $\delta_0$ in the vevs of the fields where checked in \cite{CLZ}. In this work we also take them into account. Furthermore, our aim is to consider the effect of the warping and the addition of an anti-D3 brane. To determine how much 
these corrections can change the scale hierarchy.

\subsection{More general fluxes and dilaton stabilisation effects}

Next, we consider the case of generic fluxes, but a near the conifold vacuum. The approximate value of $z_0$ is obtained by the equation
$D_z W=0$  to leading order:
\begin{eqnarray}
D_z W&=&(F_1-\tau_0 H_1)\beta \ln z+(F_1-\tau_0H_1)(\alpha+\beta)+
(F_2-\tau_0 H_2)\partial_z{\Pi^0_4}\\
&&\hskip 1cm  - (F_3-\tau_0 H_3)\partial_z{\Pi^0_1}-(F_4-\tau_0 H_4)\partial_z{\Pi^0_2}
+\partial_z K_0 W_0 ,\nn
\end{eqnarray}
where 
\begin{eqnarray}
\tau_0=\frac{F\Sigma \bar\Pi^0}{H\Sigma \bar\Pi^0}\,, \qquad \qquad 
W_0= (F-\tau_0 H)\Sigma \Pi^0 \,.\label{t0G}
\end{eqnarray}
(Note that $\tau_0$ here depends on the fluxes).
This  gives the solution for $z_0$ for any set of fluxes as:
\begin{eqnarray}
z_0&\sim& \exp\left(-\left(\frac{\alpha}{\beta}+1\right)+ \right.  \\
&& \left. \frac{-(F_2-\tau_0 H_2)\partial_z{\Pi}_4^0+(F_3-\tau_0 H_3)\partial_z{\Pi^0_1}+(F_4-\tau_0 H_4)\partial_z{\Pi^0_2}-\partial_z K_0 W_0}{\beta(F_1-\tau_0 H_1)}\right)\,.  \nn
\label{aproxG}
\end{eqnarray}
This section describes in detail the stabilization procedure for the axio-dilaton and the complex structure in Minkowski vacua of type IIB strings on one parameter Calabi-Yau manifolds.  The results are generic. In particular we also describe the stabilization of those two fields near to the conifold singularity. The results shown here, are a review of \cite{CLZ}
 but are presented in a  self contained manner,
 in order to be useful for further reference.  Next we will consider the corrections to the scalar potential arising from the warping.
 

\section{Vacua with warping corrections}    

\label{sec3}
In this Section we present the solutions for the vacua including the modification
to the K\"ahler potential given by the warping. Since the fluxes do not depend on the K\"ahler moduli, we have a no-scale structure for the scalar potential, which then simplifies to (\ref{Vnoscale}). The scalar potential now has indices which run only over the axio-dilaton and the complex structure modulus.

In \cite{CLZ} we looked for solutions to the equations $D_zW=0$ and $D_{\tau} W$, i.e. a simultaneous stabilization of the dilaton and the complex structure modulus. In \cite{Grana}, they argue that it is possible to stabilise first the dilaton and the remaining scalar potential is given by their equation (2.17):
\be\label{Vgrana}
V_G= \frac{\pi^{3/2}}{\kappa_{10}}\frac{g_s}{({\rm Im} \rho)^3}\left[c\log\frac{\Lambda_0^3}{l_s^3|z|}+ c' \frac{g_s(\alpha'M)^2}{l_s^4|z|^{4/3}}\right]^{-1}
\left|\frac{M}{2\pi i}\log\frac{\Lambda_0^3}{l_s^3 z}+i \frac{K}{g_s} \right|^2\,,
\ee 
where the second term between square brackets is the correction due to warping computed using the Klebanov-Strassler solution by Douglas et al. in \cite{DST,DT}, which we discuss later. The K\"ahler modulus is $\rho$, $\Lambda_0$ is a scale, $K$ and $M$ are the fluxes, and $z$ is the complex structure modulus. This potential has a susy minimum near the conifold (approximated by $\partial_z W=0$) at 
\be\label{sgrana}
z_{G} \simeq \frac{\Lambda_0^3}{l_s^3} {\rm exp}\left(-\frac{2\pi K}{g_s M}\right)\,.
\ee
In our notation: $H_3=K$, $F_1=M$ where $K,M$ are used in \cite{Giddings:2001yu,Grana}. As discussed in \cite{Giddings:2001yu} we also need an additional flux, $H_4$, in order to stabilise the dilaton.  In \cite{Grana} their final potential only includes $K,M$, as they assumed other fluxes stabilising $\tau$ are encoded in $g_s$, and since  $\partial_\tau W=0$, they effectively only have: 
\be
V = V_0 e^{K}K^{z\bar z}|\partial_zW|^2 = V_0 e^{K}K^{z\bar z}|(\partial_z W|^2,
\ee
where $z$ includes all complex structure moduli, although here we only have one. However for general fluxes turned on, the superpotential  $W$ \eqref{fluxW} is given by
\be
W= F_1\Pi_3+F_2\Pi_4-F_3\Pi_1-F_4\Pi_2 +\tau \left(H_3\Pi_1+H_4\Pi_2-H_1\Pi_3-H_2\Pi_4\right)\,,
\ee
where $\Pi_3$ contains the $z\log z$ term and all other periods are linear in $z$. So besides the terms with $H_3,F_1$ (i.e.~$K,M$) in \eqref{Vgrana}, there are further terms arising from the other fluxes, which will appear in the potential, essentially shifting $K, M$ in a similar expression to \eqref{Vgrana}. This however does change the solution for $z_0$. 
Thus we expect that the solution to $z$ ($s$ in \cite{Grana}) will differ from \eqref{sgrana}.  Below we check this  explicitly in numerical examples. We will see that there is a huge difference between \eqref{sgrana} and the actual solution for generic fluxes.

\subsection{Numerical examples}

Let us check a few examples to see the differences with the vacua found in \cite{Grana}. We consider  examples with generic flux configurations, some of them previously explored in \cite{CLZ}. We turn on additional fluxes, by preserving  the tadpole condition. Specifically, we can turn on $F_3,H_1,F_4,H_2$ such that their contribution to the tadpole vanishes:  $-F_3 H_1-F_4H_2=0$. 

\begin{enumerate} 

\item Taking the set of non-zero fluxes $F_1=40,H_3=16, H_4=1$ \cite{CLZ} and adding  the fluxes $F_3=5, H_1=4, F_4=-10, H_2=2$ we obtain, using the approximated solution \eqref{aproxG} $z_0=0.163842\times $ $\exp[-1.27723 i]$ and $\tau=t_1+ i t_2$ as in the mentioned reference. The result is very close to the actual full numerical solution, obtained by making zero (\ref{ecu}).  The conclusion is  that the approximation (\ref{aproxG}) is valid very near to the conifold.

\item Next we consider fluxes closer to those employed as an examples in \cite{Grana}. Concretely we take: $H_3 = 50, F_1 =10, H_4= 1$ and $F_3=5, H_1=4, F_4=-10, H_2=2$, such that $K/M=5$ as in the mentioned work. In this case, we find using our approximate solution \eqref{aproxG} $z_0=0.021\exp[-1.6252 i]$ and $\tau= -8.39999 + i \,5.79991$ (so $g_s =0.172416$).   
 However  using the  solution \eqref{zH}, we obtain  $z_0 = 7.57\times 10^{-80}\exp[0.0043 i ]$! On the other hand, if we use their approximations, that is, they take $K/M=5$, $\sqrt{g_s}M = 5, 7, 12$. Taking $K=50$ and $M=10$ as we do, then one would have $g_s =0.25,0.49,1.44 $ respectively, so we cannot use the last value of $g_s$ (we will need to take $M$ larger in this case). Using their solution \eqref{sgrana} with the first two values of $g_s$ guessed using their $z_0\sim 2.66\times 10^{-55}, 1.43\times 10^{-28}$. 
 While if we use the value of $g_s$ that we obtain from the solution, $g_s\sim 0.17$, then we recover $z_0\sim 7.36 \times 10^{-80}$.  The conclusion is that the real solution differs from the ones obtained in \cite{Grana}, and that is not possible to stabilize first the axio-dilaton and then the complex structure modulus; they have to be found as a solution to (\ref{ecu}).
 
 \item We next checked the  flux configuration: $H_3=50, F_1=10, H_4=1$ and the additional fluxes, smaller than above, $F_2=0, F_3=5, F_4=5, H_1=2, H_2=2$. The solution in this case is closer to the conifold: $z_0 = 3.12\times 10^{-7}\exp[-81.18 i]$, $\tau = -4.65+i 2.35$. Since we have the same value of the $K,M$ fluxes, solutions of \cite{Grana} will be the same, but if we use our value for $g_s$,  in \eqref{sgrana} or our \eqref{zH} we get $z_0\sim 7.75\times 10^{-33}$. Again the conclusion is that the stabilization of the axio-dilaton is connected to the stabilization of the complex structure modulus, and the approximation of separating this procedure can not be carried out. 
\end{enumerate}

The results of this subsection illustrate that we consider necessary a simultaneous stabilization of $z$ and $\tau$.  And in contrast to \cite{Grana} in most of the cases the axio-dilaton can not be stabilized first to lead an effective potential for the complex structure modulus.
\smallskip



\subsection{Strong warping correction for general fluxes}

Here we explore the issue of the strong warping correction to $G_{z\bar z}$ computed in \cite{DST,DT} using the {\bf Klebanov}-Strassler (KS) solution.  In \cite{KS}, the solution is obtained for $F_3, H_3$ fluxes, both with flux number $M$, so that the warp factor is given by $M^2$ \cite{DST}, 
\be
e^{-4A}= 2^{2/3}\frac{(g_s M \alpha')^2}{|z|^{4/3}} I(\tau),
\ee
where $z$ is the conifold modulus  and $I(\tau)$ is the integral computed in \cite{KS}. As we can see, the $K$ fluxes (or any others) do not enter in this solution and indeed the KS solution does not have the form $e^A \sim e^{-2\pi K/g_sM}$ as discussed in \cite{Giddings:2001yu}. So the KS is indeed a very different correction. The following scalar potential shows the approximation to be performed. We will consider a modification of the K\"ahler potential given by $\Delta K$, and evaluate the potential for the cs modulus $z$ to be:

\begin{eqnarray}
V(\tau_0,z)=e^{K(z,\tau_0)+\Delta K} (D_{z} W(\bar\tau_0,\bar z) D_{z_c} \bar W(\tau_0,z)).
\end{eqnarray}

\subsection{On the mirror quintic and KS}

The mirror quintic has 4 3-cycles, what can be understood from the Betti number $b_3=2+2b_{2,1}=4$. Let us denote them $A, B, A', B'$, where the periods identified with them have the behavior $A\propto z$, $B\propto zLog z + \dots$, $A', B'\propto a+ b z$. The KS solution corresponds to turning on fluxes in the $A, B$ cycles, while as noted by GKP, one needs to turn on fluxes also in the $A',B'$ cycles to stabilise the axio-dilaton, and such that $W\ne 0 $ for $\tau $ when $z=0$.   GKP work says that ``in the vicinity" if the conifold, we have the $A,B$ cycles, but they do not comment on what happens when we also add fluxes on the cycles $A',B'$ to stabilise $\tau$. Indeed,  from having only $K, M$ to having $K,M,K'$, the  solution to $z$ is modified as (their eq. (3.18) and our equation (\ref{zH}))
\be\label{zprime}
z\sim e^{\frac{2\pi K}{K'} f(0)}\,,
\ee
where $f(0)$ denotes a function of $z$ evaluated at the conifold, that is encoded in (\ref{zH}). As far as \eqref{zprime} goes, so long as $K'\gg K$, it looks like we can approach as much as we want to the conifold, just as is done for the example of the mirror quintic CY. But as is discussed in \cite{Giddings:2001yu},   Candelas-de la Ossa \cite{comments} showed that the conic coordinate $r$ goes as $r\sim z^{1/3}\sim e^{A}$, where $e^A$ is the warp factor of the D3-brane. 
Our solutions with more fluxes are indeed of the form \eqref{zprime} as we have fluxes in $A', B'$ as well as $A,B$. So, if $z\sim e^{3A}$, then in the case with more fluxes than just $K,M$ as in KS, we do have that the warp factor goes as
\be
z^{1/3}\sim e^A \sim e^{\frac{2\pi K}{K'} f(0)/3}.
\ee
Or more generally we will have the solution (\ref{aproxG}). But, $\tilde g_{mn}$ in \eqref{metric} will not be the KS metric. However we assume that the warping corrections can be employed in this setup, and we adapt the formula for the corrections to the K\"ahler potential.

\subsection{Strongly warped mirror quintic}


In order to analyze the complete scenario where the dilaton field is not a constant, we cast it by the use of $i(\tau-\overline{\tau})=1/g_s$.   The corrected K\"ahler potential is given by \eqref{kahler}, plus the strong warping correction given by \cite{DST,DT} (we use the conventions of \cite{Bento:2021nbb})
\be\label{KC2}
\Delta K(z,\bar z,\tau,\bar \tau) = \frac{\ell_s^6}{\pi||\Omega||^2V_6} \frac{9c'(g_sF_1)^2}{(2\pi)^4{\cal V}^{2/3} }\,|z|^{2/3} ,
\ee
where $c'=1.18$  (we will take $V_6=l_s^6$ later) and $||\Omega||$ is the norm of the $(3,0)$ form. On the other hand, the superpotential \eqref{fluxW} stays the same. For the near the conifold vacua we have 

\begin{eqnarray}
K(z,\bar{z},\tau,\bar{\tau})&=& -\ln \left[-i \left(\tau-\bar \tau \right)\right] -\ln~[-i\bar{\Pi}^{T} \Sigma \Pi] - 2\ln~\left[{\cal V} \right]+ \frac{c}{(\tau-\bar{\tau})^2} |z|^{2/3},  \\ \nonumber
W(z,\tau)&=& F_1\Pi_3+F_2\Pi_4-F_3\Pi_1-F_4\Pi_2 +\tau \left(H_3\Pi_1+H_4\Pi_2-H_1\Pi_3-H_2\Pi_4\right)\,,
\end{eqnarray} where we have already written $g_s$ explicitly in terms of the axio-dilaton, since we want to stabilise both $z$ and $\tau$. In the last term of $K$ we absorb all the constants in $c$. Despite the strong warping correction and due to the form of the K\"ahler potential for the K\"ahler modulus $\rho$,  the scalar potential for $\tau$ and $z$ preserves approximately the no-scale structure \cite{Blumenhagen:2019qcg, Alvarez-Garcia:2020pxd} so the potential can be written as
$$V_w=e^{K}\Big(g^{z \bar z}D_z W D_{\bar z} \overline W + g^{z \bar \tau}D_z W D_{\bar \tau} \overline W +g^{\tau \bar \tau}D_\tau W D_{\bar \tau} \overline W+ g^{\tau \bar z}D_\tau W D_{\bar z} \overline W\Big) .$$
The Minkowski minimum  of $V_w$ still satisfies $D_zW=D_\tau W=0$.  We look for solutions to Minkowski vacua  near the conifold and for several values of the fluxes. As can be expected, we have found that there is a slightly change in the value of $z$ but $\tau$ value remains basically unchanged. We can see this in the examples of vacua shown in tables \ref{tab1} and \ref{tab2}.

\begin{table}[htp]
\centering
\scalebox{1}{
\begin{tabular}{|c|c|c|c|c|c|c|} \hline
&$r_0,~r$ & $\theta_0,~\theta$& $t_1$ & $t_2$&$(F_1,H_3,H_4)$& $\vec{F}\Sigma \vec{H}$\\ \hline
1&$\begin{array}{c}3.78\times10^{-8}\\2.24\times10^{-7} \end{array}$ 
&$\begin{array}{c}-17.12 \\-15.68 \end{array}$
&$-0.89$&$1.03$&$(24, 75, 2)$ & $1800$\\
 \hline 
2&$\begin{array}{c} 3.06\times10^{-5}\\2.79\times10^{-5} \end{array}$
&$\begin{array}{c} -11.26\\-11.19 \end{array}$
&$-0.89$&$1.03$&$(12, 25, 1)$ & $300$ \\
 \hline
3 &$\begin{array}{c}4.47\times 10^{-4} \\4.67\times 10^{-4} \end{array}$
&$\begin{array}{c}-8.92\\ -8.91  \end{array}$
&$-0.37$&$0.43$ &$(5, 20, 1)$ & $100$ \\
  \hline 
4 & $\begin{array}{c}5.23\times10^{-5} \\4.97\times10^{-5} \end{array}$
&$\begin{array}{c}-10.79\\-10.42 \end{array}$
&$-1.94$&$2.23$&$(52, 48, 2)$ & $2496$ \\
\hline
5& $\begin{array}{c} 2.62\times 10^{-4}\\2.73\times 10^{-4} \end{array}$
&$\begin{array}{c}-9.31\\-9.36 \end{array}$ 
&$-2.01$ & $2.31$ &$(27,21,1)$ & $567$ \\
\hline
\end{tabular}}
\caption{Comparison of the solutions of the fields complex structure $z=re^{i\theta}$ and axio-dilaton $\tau=t_1+it_2$ in the minima of the potential $V$ and $V_w$ when only the fluxes $F_1,H_3,H_4$ are on.  $\theta_0$ and $r_0$  represent the vacua values without the strong warping correction to the K\"ahler potential.}
\label{tab1}
\end{table}
We can understand this small change, since regardless of the extra term in the covariant derivative $D_\tau W=\partial_\tau W+ K_\tau W$, with the strong warping correction, near the conifold, the additional term is tiny. One can express the new covariant derivative $D_{\tau} W'$ in terms of the non-corrected covariant derivative $D_{\tau} W$ as: $D_{\tau} W'=D_{\tau}W + const \frac{|z|^{2/3}}{(\tau-\bar\tau)}$. Hence for $|z|\ll 1$, $D_{\tau} W'\approx D_{\tau} W$, and the position of the minimum in $\tau$ for $V_w$ (and $V$) still satisfies the first equation in \eqref{tT}. One has approximately  $\tau=\frac{F \Sigma\bar \Pi}{H\Sigma \bar \Pi}$, that
for $z=0$ tends to the value without correction. This is the case for a generic flux configuration. 

To get weak coupling regime $g_s< 1$ then the condition $\Im \Big(\frac{F\Sigma \overline{\Pi}}{H \Sigma \overline{\Pi}}\Big)>1$ must be satisfied. This condition simplifies to  $\Im \Big(-\frac{F_1 \Pi_3^0}{H_4\Pi_2^0} \Big)>1$ with the minimal fluxes required for the stabilization of the dilaton and complex structure modulus near the conifold, namely with just the fluxes $F_1$, $H_3$ and $H_4$ turned on. For the mirror quintic when putting the numerical constants of the periods we obtain $F_1/H_4\gtrsim 12$. Hence, in the mirror quintic and in order to get a value of $g_s$ in the weak coupling regime with $|z|\ll 1$ and minimal configuration of fluxes, $F_1$ and $H_3$ must be of order $\mathcal{O}(10^1)$ both. The contribution to tadpole condition \cite{CLZ}  given by $F\Sigma H$ due to the fluxes will be of order $\mathcal{O}(10^2)$ (see table~\ref{tab1}). It is easier to satisfy $\Im \Big(\frac{F\Sigma \overline{\Pi}}{H \Sigma \overline{\Pi}}\Big)>1$ and $H_3\gg H_4$ to achieve $g_s<1$ and as well to be near the conifold when all the fluxes are on. In these cases, the contribution to the tadpole can be made small with an interplay between the value and signs of the fluxes (See example in table \ref{tab2}).

On the other hand, as we will see in the next section, the addition of an $\overline{D}3$-brane does not necessarily destabilize the complex structure modulus if $\sqrt{g_s}F_1 \lesssim 7$ as was established in \cite{Grana}. We have found examples where $F_1$ is small (despite $g_s<1$) and $\tau$ and $z$ are stabilized. The value of $F_1$ is not the only which determines the stabilization of $z$. The stabilization/destabilization of the conifold modulus depends on the values of all the fluxes. 

The mass hierarchies between the dilaton and $z$ is well determined. That is, the dilaton mass is lighter (or same order of magnitude) than the complex structure in most of the flux configurations we have studied. We can understand this behavior from estimating  the masses of the scalar field near the conifold as we {\bf obtain} next. 
\begin{table}[htp]
\centering
\scalebox{0.75}{
\begin{tabular}{|c|c|c|c|c|c|c|} \hline
$r_0,~r$ & $\theta_0,~\theta$& $t_1$ & $t_2$& $\vec{F},~\vec{H}$ & Masses &$\vec{F}\Sigma \vec{H}$ \\ \hline
$\begin{array}{c} 7.95\times 10^{-6} \\8.22\times 10^{-6}\end{array}$
& $\begin{array}{c}-32.84\\-32.90 \end{array}$
& $-4.36$& $3.58$& $\begin{array}{c}(20, 6,5,-5)\\ (2,2,30,1)\end{array}$& 
$\begin{array}{c}
m^{2}_{o_i}:(1.5\times10^{5}, 1.5\times10^{5},2.5\times10^{-5}, 2.5\times10^{-5}) \\
m_i^2:(0.51, 0.66, 2.5\times10^{-5}, 2.5\times10^{-5}) \\
\end{array}$& $606$
\\ \hline
$\begin{array}{c}4.21\times10^{-7}\\ 4.27\times10^{-7} \end{array}$ 
&$\begin{array}{c}-9.89\\-9.86\end{array}$ 
&  $0.19$& $5.22$ & 
$\begin{array}{c}(9, 10, -30, -4)\\ (3, 5, 10, 0)\end{array}$
&$\begin{array}{c}
m^2_{o_i}:(9\times10^{6}, 9\times10^{6}, 1.7\times10^{-5}, 1.7\times10^{-5}) \\
m^2_i:(0.22, 0.21, 1.7\times10^{-5}, 1.7\times10^{-5}) \\
\end{array}
$ & $200$
\\ \hline
 $\begin{array}{c}1.71\times 10^{-5}\\ 1.70\times 10^{-5}\end{array}$
& $-106.05$ & $-20.33$&$13.16$
&$\begin{array}{c}(24,17,13,-29)\\ (3,3,66,1)\end{array}$& 
$\begin{array}{c}
m^2_{o_i}:(98949, 98936, 1.1\times10^{-4},  1.01\times10^{-4})\\
m^2_{i}:(21.25, 21.06, 1.1\times10^{-4}, 1.1\times10^{-4}) 
\end{array}$ & $1649$
\\ \hline
$\begin{array}{c}7.37\times10^{-8}\\ 1.78\times10^{-7}\end{array}$
&$\begin{array}{c}-27.67 \\-27.25 \end{array}$
& $2.66$& $1.32$& $\begin{array}{c}(20, 10, 10, 0)\\ (10, 0, 30, 1)\end{array}$ &
$\begin{array}{c}
m^2_{o_i}:(6.7\times10^{8}, 6.7\times10^{8}, 5.9\times10^{-6}, 5.9\times10^{-6})\\
m_i:(0.0002, 9\times10^{-5}, 5.9\times10^{-6}, 5.9\times10^{-6}) \\
\end{array} 
$& $510$
\\ \hline
$\begin{array}{c} 1.91\times10^{-19}\\2.97\times10^{-15} \end{array}$ 
&$\begin{array}{c} 50.86 \\51.65  \end{array}$  
& $6.97$ & $4.28$  &$\begin{array}{c}(3, 9, 7, -6)\\(0, 5, 5, -1)\end{array}$& 
$ \begin{array}{c}
m^2_{o_i}:(2.2\times10^{29}, 2.2\times10^{29}, 1.0\times10^{-5}, 1.0\times10^{-5})\\
m_i:(0.00011, 0.000017, 1.0\times10^{-5}, 1.0\times10^{-5}) \\
\end{array}$& $36$
\\
\hline
\end{tabular}} 
\caption{  We show examples of Minkowski vacua for different configurations of fluxes. In the sixth column we show the comparison of the numerical mass matrix eigenvalues for the potential $V$ and $V_w$. The value of the volume is fixed to be $\mathcal{V}=10^3$ where our numerical results show larger effects of the warping correction. This voume warranties the supergravity approximation, giving distances much larger than $l_s$.}
\label{tab2} 
\end{table}
\subsection{Masses calculation}

In this Section we obtain explicit formulas for the masses of the moduli, for 
a near the conifold vacuum. We also explicitly check a fact connected with the impossibility of stabilizing first the axio-dilaton, that is because generically $z$ is heavier than $\tau$.

The square masses of the canonical normalized fields are given by the eigenvalues of the matrix $\big(M^{I}_{J}\big)^2=\frac{1}{2}G^{IK}V_{KJ}$
\begin{equation}\label{MassM}
\big(M_{J}^{I}\big)^2=\frac{1}{2}\begin{pmatrix}
         g^{i \overline {k}}V_{\overline k j}~ &~ g^{i\overline{k} }V_{\overline{k}\overline{j}} \\
         g^{\overline{j}k}V_{k i}~&~ g^{\overline{j}k}V_{k \overline{i}}  
       \end{pmatrix},
\end{equation}
where in the minimum $D_iW=0$ and  
$$V_{JK}=\partial_K \partial_J V =e^{K}\Big(g^{i \overline{j}}(\partial_JD_iW) (\partial_K D_{\overline j} \overline W )+ ( J \leftrightarrow K) \Big).$$

In the limit of weak warping or without the correction \eqref{KC2} and near the conifold the mass matrix in  eq.~\eqref{MassM} is in good approximation diagonal and eigenvalues are degenerate. The two eigenvalues are $\lambda_{o_{z}}=\frac{1}{2}g^{z\overline z}V_{z\overline z}$ and $\lambda _{o_{\tau}}=\frac{1}{2}g^{\tau \overline \tau} V_{\tau \overline \tau}$. In those limits the two masses are then, 

\begin{subequations}
 \begin{align}
m_{o_z}^2 &\sim \frac{e^{K_0}}{2\mathcal{V}^2}(g^{z\overline z}_0)^2
\Big ( \frac{|\Pi_3^{0}|^2|F_1-\tau H_1|^2}{|z|^2}\Big),\label{massDiv} \\
m_{o_\tau}^2 &\sim \frac{e^{K_0}}{2\mathcal{V}^2} \Big(\big|W_0\big|^2+\big|\bar{\Pi}_4^0(F_2-\tau H_2)
+ \bar{\Pi}_3^0(F_1-\tau H_1)-\bar{\Pi}_2^0(F_4- \tau H_4)\big|^2 \Big),
\end{align}\label{MassesSC}
\end{subequations} where $e^{K_0}=\lim_{_{z\rightarrow 0}}e^{K+\ln(\mathcal{V}^2)}$ and $g_{0}^{i\bar i}= \lim_{_{i\rightarrow 0}} g^{i \bar i}$. $W_0=\lim_{z\rightarrow 0} W$. Clearly, near the conifold where $|z|\ll 1$, the masses have the hierarchy $m_z\gg m_{\tau}$ found before \cite{Blumenhagen:2019qcg, Alvarez-Garcia:2020pxd, Seo:2021kyi,CLZ}. We can observe this hierarchy in all examples examined near the conifold (see table \ref{tab2} where we put the exact numerical eigenvalues).     

With the K\"ahler potential correction \eqref{KC2}, the K\"ahler metric is now non-diagonal and, each entry is dependent on all moduli. The components of the metric are  
\begin{align}
g_{z\bar{z}}= & g_{z \overline z}^{(z)}+\frac{1}{9(\tau- \overline {\tau})^2}\frac{c}{|z|^{4/3}},
 & & g_{z \overline {\tau}}= \frac{2c}{3(\tau-\overline{\tau})^3}\frac{\overline z^{1/3}}{z^{2/3}}, & \nonumber
\\
\nonumber
g_{\tau \overline{\tau}}=&-\frac{1}{(\tau- \overline \tau)^2}- \frac{6c |z|^{2/3}}{(\tau-\overline{\tau})^4}, 
 & & g_{\tau \overline z}= -\frac{2c}{3(\tau-\overline{\tau})^3}\frac{z^{1/3}}{\overline z^{2/3}}, & \nonumber  
\end{align}
where we named $g_{z \overline{z}}^{(z)}$ to the $z \overline{z}$ metric component obtained from \eqref{kahler}. We can note that if $\mathcal{V}\gg 0$ ($c\rightarrow 0$) this  reduces to the original metric. More precisely the strong warped regime is that where $\mathcal{V}|z|^2 \ll 1$ (or $\mathcal{V}^{2/3}|z|^{4/3} \ll 1$ ) \cite{Blumenhagen:2019qcg}. 

The exact eigenvalues of the matrix $\big(M_{J}^{I}\big)^2$ with the non-diagonal metric are harder to calculate but near the conifold (as in the case without strong warping correction) some of the terms appearing in $\big(M_{I}^{J}\big)^2$are sub-leading.
Hence, near the conifold, the mass matrix is approximately   
\begin{equation}
(M_{J}^{I})^2 \sim \begin{pmatrix}
         M^{z}_z ~&0~&M_{\bar{z}}^{z}~&0 \\
        M^{\tau}_z~&~M^{\tau}_\tau &~M^{\tau}_{\bar z} &M^{\tau}_{\bar \tau} \\
        M^{\bar z}_{z}~&~0 &~ M^{\bar z}_{\bar z}&0\\
        M^{\bar \tau}_{\bar z}~&~M^{\bar \tau}_{\tau} &~ M^{\bar \tau}_{\bar z}~&~ M_{\bar \tau}^{\bar \tau}\\
       \end{pmatrix}.
       \label{mmatrix}
\end{equation}
The four eigenvalues of a matrix as the previous one are easy to calculate. Eigenvalues are given by $\lambda_{1,2}=M_{z}^{z}\pm \sqrt{M_{z}^{\overline z}M_{\overline z}^{z}}$ and  $\lambda_{3,4}=M_{\tau}^{\tau}\pm \sqrt{M_{\tau}^{\overline \tau}M_{\overline \tau}^{\tau}}$, and hence the square masses of the four real fields to dominant order are roughly

\begin{subequations}
\begin{align}
m_1^2 &\sim \frac{e^{K_0}}{\mathcal{V}^2}
\Big (9 |W_0|^2+
c_1 \frac{\mathcal{V}^{4/3}(\tau-\bar{\tau})^4|\Pi_3^0|^2|F_1-H_1\tau|^2|z|^{2/3}}{F_1^4}\Big),\\
m_{2}^2 &\sim  \frac{e^{K_0}}{\mathcal{V}^2}
\Big (|W_0|^2+
c_1\frac{\mathcal{V}^{4/3}(\tau-\bar{\tau})^4|\Pi_3^0|^2|F_1-H_1\tau|^2|z|^{2/3}}{F_1^4} \Big),  \\
m_3^2 &\sim \frac{e^{K_0}}{\mathcal{V}^2}\Big(\big|W_0\big|^2+\big|\bar{\Pi}_4^0(F_2-\tau H_2)
+ \bar{\Pi}_3^0(F_1-\tau H_1)-\bar{\Pi}_2^0(F_4- \tau H_4)\big|^2 \Big), \\ 
m_{4}^2 &\sim m_{3}^2,
\end{align}\label{MassesCC}
\end{subequations}
where   
$c_1= \frac{16\pi^2 ||\Omega||^4V_6^2 (2\pi)^8}{c'^2 \ell_s^{12}}$. Near the conifold (in particular for the 
mirror quintic CY) $W_0$ is given by
\begin{equation}
W_0= \big|\Pi_2^0(F_4-\tau H_4)- \Pi_3^0(F_1-\tau H_1)- \Pi_4^0 (F_2-\tau H_2)\big|^2. \nn
\end{equation} We can see in eqs.~\eqref{MassesSC} that masses depends on the value $W_0$ and hence in the values of the fluxes. With all fluxes on, the value of $W_0$ is difficult to adjust it small but in the mass it is suppressed by the volume. If the volume is not too large ($\mathcal{V}< 10$) it can be comparable to the other terms in $m_1$ and $m_2$. 

Table \ref{tab2} shows the mass eigenvalues in both cases, with and without the warping correction. The values shown are the exact eigenvalues of the mass matrix \eqref{MassM}. These eigenvalues are well estimated by  eqs.~\eqref{MassesCC} and \eqref{MassesSC}. 
As can be observed from these expressions, the masses of the two $\tau$ real fields are approximately the same despite adding the correction to the Kähler potential. For the heavier eigenvalues without the warping correction, one obtains $m^2_{O_z} \sim 1/|z|^2$, as can be seen in (\ref{massDiv}). Thus, in the uncorrected theory, there are fields with unbounded masses, primarily governed by the complex structure. The warping correction acts as a form of regularization for these masses, eliminating their singularity at $z=0$. This regularization is evident in the asymptotic behavior near the conifold once the warping corrections are incorporated. This phenomenon is also illustrated in Figure 1, where one can observe that the scalar potential becomes smoother at the minimum when Kähler corrections are taken into account.  Let us point out that for the uncorrected theory, the values of the mass eigenvalues will give rise to modes orders of magnitude heavier than the Planck scale. Thus, the effective theory constructed with the string
zero modes will no longer be a good approximation, as heavier string excitations have been excluded in our description. 
Thus it becomes necessary to incorporate
warping corrections, to achieve a well behaved, non singular theory.

The situation remains the same when the potential of the $\overline{D3}$-brane is added as we observe later. However, near the conifold, the masses of the $z$ fields ($r$ and $\theta$ {\color{blue}$p$} in the polar basis) are notably diminished when we consider the warping correction. With the strong warping correction and $\mathcal{V}\gg 1$ we have $m_1> m_{2}> m_{3,4}$.  The axio-dilaton is always the lighter field and it seems difficult to invert that hierarchy while keeping $g_s<1$ and $|z|<1$.  From the analytical  and numerical explorations we conclude, that there is a clear hierarchy between the mass of the cs and the one of the axio-dilaton. So, it is not possible to stabilize first $\tau$ and analyze an effective potential for $z$.


\section{$\overline{D3}$-brane Uplifting of the potential}  

\label{sec4}
 We are now ready to check the behaviour of the minimum for all the fields when adding the uplifting potential due to the addition of an anti-$D3$-brane\footnote{This can also be done within linear supergravity by introducing a nilpotent superfield \cite{Kallosh:2014wsa,Bergshoeff:2015jxa}. }.  In this Section we add this correction to the scalar potential corrected by the warping, which was previously obtained. This contribution is given by (see \cite{Grana,Bento:2021nbb} for derivation and discussion)
 
 
 \be
 V_{\overline{D3}} =\frac{(2\pi)^4}{{8 \pi \cal V}^{4/3}} \frac{c'' \,|z|^{4/3}}{(g_sF_1)^2} \,M_{Pl}^4  = d \,|z|^{4/3} (\tau-\bar \tau)^2\,,
 \ee
where $c'' = 1.75$ and in the last equality we wrote $g_s$ in terms of the axio-dilaton as we did before. We now can look for minima of the potential computed in the previous sections, plus the anti-$D3$ contribution above, this is we will search for minima of the corrected potential
\begin{equation}
    V_{tot}=e^{K}(g^{a\bar b}D_a W D_{\bar b} \overline W)+ V_{\overline{D3}}.
\end{equation}
 with $a,b=\{z,\tau \}$ as before.
 \subsection{Criteria for finding vacua}
 
According to the joint analysis of the stabilization of the axio-dilaton and the complex structure modulus, near to the conifold, the stabilization value of the axio-dilaton is the same despite the inclusion of the correction and the $\overline{D3}$. Then, we can minimize $V_{tot}$ with respect to $z$ setting the value of $\tau$ as a constant. Following \cite{Grana}, with the solution to $\partial_z V_{tot}=0$ we can get a lower bound for not losing the minimum when we add the $\overline{D}3$-brane. When all fluxes are included, however, the situation is quite different and, such a bound is impossible to get analytically because the associated equations for $r$ and $\theta$ are transcendental equations. 

The total $z$ potential is
\begin{equation}
V_{tot}(z,\overline{z})\approx\frac{|z|^{4/3}}{g_sc_v\mathcal{V}^{4/3}}\Big(9e^{K^{(z)}_0}g^{\bar{z}{z}}D_{z}W D_{\overline z}W + \frac{c_v d_v}{g_s} \Big), 
\end{equation}
with $d_v=(2\pi)^4c''/F_1^2$ and $c_v=9c'F_1^2/(2\pi)^4\pi\|\Omega\|^2$ ($l_s=1$, $V_6=1$). The $z$ derivative include the terms $\partial_z D_z W=\partial^2_{z}W+K_{zz}W+K_{z}\partial_{z}W$ and $\partial_z D_{\overline z} W=K_{z\overline z}W$. If $z$ is close to the conifold and $W$ includes all fluxes with general expansions of the periods to order more than one, then the constant terms in $W$ (or derivatives of) with $K$'s derivatives could be the dominant ones. However, making the assumption that $D_z W\sim \partial_z W$, $D_{\overline z}\overline W \sim \partial_{\overline z} \overline {W}$ that is, if we consider just linear terms in the periods (and hence in the superpotential) as in \cite{Grana}, the optimization can be performed. Solving $\partial_z V_{tot}=0$ for $r$ and $p$ gives:

\begin{eqnarray}
r&=& \exp\Big(-\frac{9}{12}-\frac{\overline h}{2(F_1-\overline\tau H_1)\overline{\beta}}-\frac{h}{2(F_1-\tau H_1)\beta}\pm \sqrt{-\frac{c_vd_ve^{-K_0^{(z)}}}{9g_s|F_1-\tau H_1|^2|\beta|^2}+\frac{9}{16}}
 \Big), \nn \\
\theta&=&\frac{i}{2}\Big(\frac{h}{(F_1-\tau H_1)\beta}-\frac{\overline h}{(F_1-\overline \tau H_1)\overline \beta}\Big),
\end{eqnarray}
where $h$ is the constant term in $\partial_z W$. In the quintic it is $h=(F_1-\tau H_1)(\alpha+\beta)+(F_2-\tau H_2)\Pi_{4}^1-(F_3-\tau H_3)\Pi_1^1-(F_4-\tau H_4)\Pi_2^1$.
With this solution, in order to get two different values of $r$ (a maximum and a minimum), the next bound must be satisfied, 
 \begin{equation}
  g_s|F_1-\tau H_1|^2> \frac{2c'c''e^{-K^{(z)}_0}}{9\pi |\beta|^2}.
 \end{equation}
 
In the mirror quintic $e^{-K^{(z)}_0}$ is slightly different from 1 ($\sim 1.11$) and then we get $\sqrt{g_s|F_1-\tau H_1|^2}$$\gtrsim \sqrt{51}$ in order to prevent the destabilization of the local minimum. It is important to note that this bound is still very imprecise and does not apply at all in our case but could be a guide towards avoiding the destabilization of the conifold modulus.  

On the other hand, a general bound (if exists) should include all the fluxes in a different expression, making the possibility of having solutions with small tadpole satisfying that bound. Tables \ref{tab3}, \ref{tab4} shows examples with a small tadpole contribution of the fluxes and a local minimum of the potential. 
\begin{table}[htp]
\centering\small
\scalebox{1}{
\begin{tabular}{|p{1.65mm}|c|c|c|c|c|c|c|c|c|c|} \hline
&$r_0,~r,~r_f$ & $\theta_0,~\theta,~\theta_f$& $t_1$ & $t_2$& $\vec{F},~\vec{H}$ &$\vec{F}\Sigma \vec{H}$ \\ \hline
1& $\begin{array}{c}5.90\times10^{-5} \\5.86\times10^{-5} \\3.11\times10^{-5} \end{array}$ &$\begin{array}{c} -9.48\\ -9.48 \\ -9.46 \end{array}$   
&$\begin{array}{c} \\ 0.77\\ 0.81 \end{array}$
&$\begin{array}{c} \\ 4.40 \\ 3.39 \end{array}$
&$\begin{array}{c}(13, -13, -1, 4)\\ (3, -7, 9, 1)\end{array}$ 
& $135$
\\ \hline
2&$\begin{array}{c} 2.72\times10^{-3}\\2.34\times10^{-3}\\2.22\times 10^{-3} \end{array}$ &
$\begin{array}{c} -23.90\\ -23.76 \\ -23.77 \end{array}$&
$-2.03$&
$1.03$ &
$\begin{array}{c}(22,17,12,-10)\\ (1,6,40,4)\end{array}$
&$996$
\\ \hline
3&$\begin{array}{c}8.26\times10^{-6}\\ 8.08\times10^{-6} \\ 4.83\times10^{-6} \end{array}$ 
&$\begin{array}{c}21.04\\ 21.09 \\ 21.07 \end{array}$
& $0.61$ & $2.51$ &
$\begin{array}{c}(12, 4, -25, 2)\\ (-1, -4, 15, 1)\end{array}$
& $167$
\\ \hline
4& $\begin{array}{c}2.09\times 10^{-4}\\2.07\times 10^{-4}\\1.92\times 10^{-4}\end{array}$
&$\begin{array}{c}-57.10\\-57.13\\-57.13\end{array}$
& $-3.83$&$2.77$&
$\begin{array}{c}(14,20,28,-13)\\ (4,7,58,2)\end{array}$
& $831$
\\ \hline
5 &$\begin{array}{c}4.82\times 10^{-9}\\ 7.5\times 10^{-9}\\4.34\times 10^{-9}\end{array}$
& $\begin{array}{c} -69.74\\-69.64\\-69.58 \end{array}$
&$-6.40$&$2.18$&
$\begin{array}{c}(10,10,15,4)\\(0,-5,15,0)\end{array}$
& $170$
\\ \hline
6 &$\begin{array}{c} 8.6\times10^{-7} \\ 8.15\times10^{-7} \\ 5.1\times10^{-7} \end{array}$  
& $\begin{array}{c} -24.82  \\ -24.88 \\ -24.85  \end{array}$
&$-2.26$&$3.01$& 
$\begin{array}{c}(12, 15, 12, -2)\\ (1, -2, 17, 1)\end{array}$
& $203$
\\ \hline
7&$\begin{array}{c}1.04\times10^{-13}\\ 1.76\times10^{-11}\\ 1.14\times10^{-11}\end{array}$& 
$\begin{array}{c}-6.74\\-6.85 \\ -6.85
\end{array}$
& $-1.61$ & $1.82$  &$\begin{array}{c}(5, -6, 2, 4)\\ (-1, -9, 7, 0)\end{array}$
& $73$
\\ \hline
\end{tabular}}
\caption{Some examples of vacua for the potential $V$ and $V_w$ and $V_{tot}$ for arbitrary configurations of fluxes. In all cases $\mathcal{V}=10^3$. The first columns show the minimum solutions for the real fields $r$, $\theta$ in the three cases. The values of $t_1$ and $t_2$ does not suffer any chage except between $V_w$ and $V_{tot}$ in the example number 1.}
\label{tab3}
\end{table}

\begin{figure}[ht]
    \centering
    \includegraphics[scale=0.3]{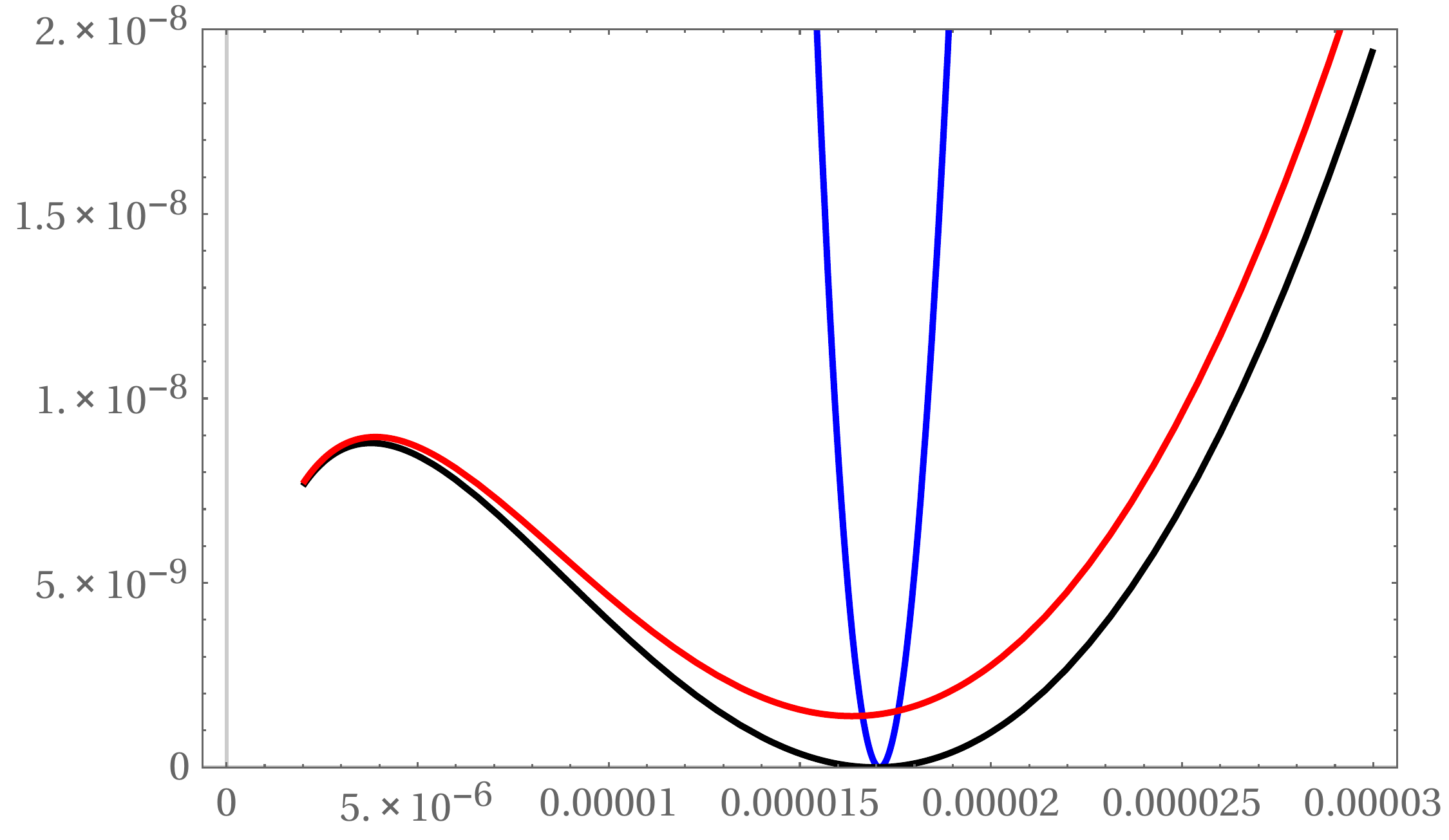}
    \caption{The potentials $V$, $V_w$ and $V_{tot}$ for the $r$ direction of the complex structure modulus $z$. The blue line corresponds to the original potential $V$, while the red and black lines  show the strong warping corrected potential $V_w$ and the potential with the addition of the $\overline{D}_3$ denoted $V_{tot}$. Graphs were drawn with $F=(24,17,13,-29)$, $H=(3,3,66,1) $ and $\mathcal{V}=10^3$. }
    \label{fig:1}
\end{figure}

To conclude this section we show in figure \ref{fig:1} the computation of the scalar potential for the three different cases. This is, we show the initial scalar potential for $z$ and $\tau$ without warping corrections, with warping corrections, and with the addition of and anti-D3 brane.

\begin{table}[htp]
\centering\small
\scalebox{0.8}{
\begin{tabular}{|c|c|c|c|c|c|c|} \hline
$r,~r_f$ & $\theta,~\theta_f$& $t_1$ & $t_2$&$(F_1,H_3,H_4)$ & $\vec{F}\Sigma \vec{H}$&Masses\\ \hline
$\begin{array}{c}9.84\times10^{-9} \\9.20\times10^{-9}\end{array}$ 
&$\begin{array}{c} -20.44 \\ -20.43\end{array}$
&$-1.12$&$1.28$& $(15, 42,1)$ & $630$ &
$\begin{array}{c}
m^2_i:(0.00014, 0.00004, 7.1\times10^{-6}, 7.1\times10^{-6})  \\
m^2_{f_i}:(0.00014, 0.000039, 7.1\times10^{-6}, 7.1\times10^{-6})
\end{array}$
\\ \hline 
$\begin{array}{c} 3.82\times10^{-7}\\3.67\times10^{-7} \end{array}$ 
&$\begin{array}{c}-15.32 \\-15.32 \end{array}$
&$-2.76$&$3.17$& $(37, 34, 1)$ & $1258$ &
$\begin{array}{c}
m^2_i:(0.0015, 0.00089, 0.000018, 0.000018) \\ 
m^2_{f_i}:(0.0014, 0.00082, 0.000018, 0.000018)
\end{array}$
\\ \hline
$\begin{array}{c} 1.07\times10^{-9}\\6.06\times10^{-10} \end{array}$
& $\begin{array}{c} -21.66\\-21.69 \end{array}$ 
&$-0.075$ & $0.086$ &$(1, 50, 1)$ & $50$ &
$\begin{array}{c}
m^2_{i}:(7\times10^{-6}, 1.6\times10^{-6} , 4.7\times10^{-7}, 4.7\times10^{-7}) \\
m^2_{f_i}:(5.1\times10^{-6}, 5.1\times10^{-7}, 4.7\times10^{-7}, 4.7\times10^{-7})
\end{array}$ \\ \hline
 $\begin{array}{c}6.24\times10^{-7}\\ 5.85\times10^{-7}\end{array}$
&$\begin{array}{c} -15.00\\ -15.00\end{array}$ 
& $-1.71$ & $1.97$ &$(23, 33, 1)$& $759$ 
& $\begin{array}{c}
m^2_i:(0.0011, 0.00075, 1.1\times10^{-5}, 1.1\times10^{-5})\\
m^2_{f_i}:(0.0011, 0.00066, 1.1\times10^{-5}, 1.1\times10^{-5})
\end{array}$ \\ \hline
$ \begin{array}{c}1.55\times10^{-5}\\1.45\times10^{-5}  \end{array}$
& $\begin{array}{c} -11.71\\ -11.71\end{array}$ &$-2.46$ & $2.82$ &$(33,26,1)$
& $858$  &
$\begin{array}{c}
m^2_i:(0.0064, 0.0052, 0.000016, 0.000016) \\
m^2_{f_i}:(0.0056, 0.0050, 0.000016, 0.000016)
\end{array}$ \\ \hline
\end{tabular}}
\caption{Vacua for the case of non zero fluxes $F_1,H_3,H_4$ . We give the mass matrix eigenvalues for Minkowski vacua, for the potential $V_w$ and $V_{tot}$ and adimensional volume $\mathcal{V}=10^3$, this is compatible with the supergravity approximation.  After the correction is implemented, also the hierarchy in general holds, although the scale
difference diminishes.}
\label{tab4}
\end{table}

\section{Conclusions and Outlook}
\label{sec5}
In this work we have explored type IIB flux orientifold compactifications on compact Calabi-Yau varieties
with one complex structure parameter, and with important warping. Following on previous results for axio-dilaton and complex structure estabilization for the non-compact case, where an instability near the conifold arises \cite{Grana}, we extend this results to the compact case finding new distinct features.
For the nummerics we concentrate in the case of the Mirror quintic CY 3-fold, with one complex structure modulus and 101 K\"ahler moduli  \cite{CLZ}, but one should expect similar results in
other one-parameter models, as the mechanism for complex structure and axio-dilaton
stabilization are identical, such that our formulae apply to find their fixed values.

In the mentioned global model we explore the implications for vacua of the previously explored correction to the  K\"ahler potential due to the warping near the conifold.  We have encountered interesting facts. First the axio-dilaton
and complex structure modulus have to be stabilized simultaneously, it is not exact to assume estabilization of the axio-dilaton first and the complex structure in a second step.  This occurs because generically the complex structure is heavier than the axio-dilaton. We have obtained analytical formulas and checked this statement in several numerical examples. Second,
it is possible to obtain solutions that are near the conifold and possess certain hierarchies between
the internal 6D scale and the space-time scale, in the compact case, and this conclusion is preserved by adding the warping corrections. With some assumptions we find a criteria to find minima and maxima of the potential. However, the assumptions are strong, such that in this compact case there is not a clear bound for the fluxes
which leads to an instability, in contrast to \cite{Grana}. We also make explicit here, an observation which is clear from previous studies. This is that the non-supersymmetric Minkowski vacua in the axio-dilaton and cs are much more dense than  supersymmetric Minkowski vacua. This occurs because to find the first we need to solve a complex equation for a complex parameter $z_0$; and to find the second implies that the same number $z_0$ fulfills two complex equations. Which would seem as an improbable destiny, but indeed these classes of vacua can be found by considering the modularity of Calabi-Yau manifolds \cite{Kachru:2020sio,Candelas:2023yrg}. The main conclusion here is that type IIB compactifications on CY with a single cs modulus have stable vacua on the cs and axio-dilaton sector. This has been tested by addding the warping corrections to the K\"ahler potential, and by adding anti-D3 branes to implement the uplifting.

These results have to be contrasted with a
work that shows that the 10D equations of motion, give some constraints for the warping \cite{Lust:2022xoq}. We plan to implement this into the compact 
internal dimensions in future work.  Additionally an statistical search for solutions has to be performed: we plan to carry out this exploration employing  machine learning techniques \cite{CaboBizet:2020cse}. Also, as the most important task, we need to address the stabilization of the K\"ahler moduli, for this aim it could be useful to select CY varieties with few parameters as the ones constructed in \cite{Braun:2009qy,Braun:2015jdy,Candelas:2016fdy}. There are currently interesting works carried out in the same family of models that can be interesting to contrast with \cite{Plauschinn:2023hjw,Cota:2023uir,Bastian:2023shf,Cicoli:2023opf,Alvarez-Garcia:2020pxd,Hebecker:2018fln} in future explorations.

\newpage 
\section*{Acknowledgements} 

We thank Ivonne Zavala for early collaboration and we appreciate her comments and suggestions. We thank useful discussions and comments from Dami\'an Mayorga Peña, Alejandro Cabo Montes de Oca, Xenia de la Ossa and   Albrecht Klemm.  YOT thanks the support of PRODEP-SEP Fellowship 2020, project {\it ''Las conjeturas del pantano en teoría de cuerdas y sus implicaciones cosmológicas"}. NGCB acknowledges: Project CONAHCyT A1-S-37752;  University of Guanajuato DAIP Projects CIIC 251/2024, CIIC 224/2023, CIIC  264/2022 and CIIC 148/2021;  the Institute of Theoretical Physics, ETH Z\"urich; the support from the ICTP through the Associates Programme (2023-2029); and the Isaac Newton Institute for Mathematical Sciences, Cambridge, 
for support and hospitality during the programme "Black holes: bridges between number theory and holographic quantum information" where work on this paper was undertaken.  This work was supported by EPSRC grant no EP/R014604/1. OLB thanks the support of CONAHCyT Ciencia de Frontera 2023 and University of Guanajuato DAIP CIIC 2021 y CIIC 2024.  

\section*{Data availability}

The data for the numerical results obtained in this paper are explicitly given in the Tables. The computer programs employed can be obtained by a request to the authors.

\begin{appendices}

\newpage

\section{Mass estimations}
\label{appA}

In this appendix we show an estimation of the moduli mass matrix by considering a K\"ahler potential $\cal{K}$ of the form
\begin{equation}
{\cal{K}}(z,\bar{z},\tau,\bar\tau)= K^0(z,\bar{z}) +K^0(\tau,\bar\tau) + K(z,\bar{z},\tau,\bar{\tau}).
\end{equation}
Near the conifold, at leading order in $|z|$, the K\"ahler potential $K^0(z,\bar{z})$ is approximated to
\begin{eqnarray}
K^0(z,\bar{z})&=&\ln \left(|z|^2\ln |z|^2 + A(z+\bar{z})+B+\dots\right),
\label{K0z}
\end{eqnarray}
with
\begin{equation}
K^0(\tau,\bar{\tau})=-i\log\left(i(\tau-\bar{\tau})\right),\qquad \text{and}
\qquad K(z,\bar{z},\tau,\bar{\tau})=\frac{c |z|^{2/3}}{(\tau-\bar\tau)^2},
\end{equation}
with $z$ the complex structure and $\tau$ the complex axio-dilaton. For a no-scale model with a superpotential given by Eq.(\ref{fluxW}), the conditions $D_zW= D_\tau W=0$ are sufficient to fix a critical point $\partial_I V=0$ where $I= (i,\bar{i})$ with $i=z,\tau$ and $V$ is the scalar potential
\begin{equation}
V=e^{\cal K} D_i W D_{\bar{j}}\bar{W} {\mathcal K}^{i\bar{j}},
\end{equation}
where ${\mathcal K}^{i\bar{j}}$ is the inverse matrix of $\mathcal{K}_{i\bar{j}}=\partial_{i\bar{j}}\mathcal{K}$.  A minimum is reached if the eigenvalues of the Hessian matrix 
\begin{equation}
\partial_{IJ} V=V_{IJ}=
\begin{pmatrix}
V_{ij}&V_{i\bar{j}}\\
V_{\bar{i}j}&V_{\bar{i}\bar{j}}
\end{pmatrix}
\end{equation}
are all positive, or equivalently, if the canonical normalized mass matrix $(M^I_J)^2$ has positive eigenvalues, with
\begin{equation}
(M^I_J)^2=\frac{1}{2}\mathcal{K}^{IM}V_{MJ}=
\begin{pmatrix}
(M^i_j)^2&(M^i_{\bar{j}})^2\\
(M^{\bar{i}}_j)^2& (M^{\bar{i}}_{\bar{j}})^2
\end{pmatrix}.
\end{equation}
At the critical point, $V_{IJ}$ is given by 
\begin{equation}
V_{IJ}=e^{\cal K}\left[ (\partial_J D_i W)(\partial_I D_{\bar{j}}\bar{W})+(\partial_ID_iW)(\partial_J D_{\bar{j}}\bar{W})\right]\mathcal{K}^{i\bar{j}}.\nonumber
\end{equation}
With the purpose to estimate the mass values near the conifold, it will be useful to denote the mass matrix as
\begin{eqnarray}
(M^I_J)^2&=&\frac{e^\mathcal{K}}{2}\mathcal{K}^{IM} E_{MP}\mathcal{K}^{PQ} E^T_{JQ},
\end{eqnarray}
where
\begin{eqnarray}
E_{MN}&=&(
e_{Mn},
e_{M\bar{n}})=
\begin{pmatrix}
e_{mn}& e_{m\bar{n}}\\
e_{\bar{m}n}&e_{\bar{m}\bar{n}}
\end{pmatrix},
\end{eqnarray}
\begin{eqnarray}
\mathcal{K}^{MN}&=&\begin{pmatrix}
0&\mathcal{K}^{m\bar{n}}\\
\mathcal{K}^{\bar{m}n}&0
\end{pmatrix}
\end{eqnarray}
and
\begin{eqnarray}
e_{Mn}&=& \partial_M D_n W,\nonumber\\
e_{M\bar{n}}&=& \partial_M D_{\bar{n}}\bar{W}.
\end{eqnarray}
 Also observe that
\begin{eqnarray}
e_{ij}&=&\left(\partial_{ij}+K^0_{ij}+K_{ij} +K_j \partial_i\right)W,\nonumber\\
e_{\bar{i}\bar{j}}&=&\left(\partial_{\bar{i}\bar{j}}+K^0_{\bar{i}\bar{j}}+K_{\bar{i}\bar{j}}+K_{\bar{j}}\partial_{\bar{i}}\right)\bar{W},\nonumber\\
e_{\bar{i}j}&=&\left(K^0_{\bar{i}j}+K_{\bar{i}j}\right)W,\nonumber\\
e_{i\bar{j}}&=&\left(K^0_{i\bar{j}}+K_{i\bar{j}}\right)\bar{W}.
\end{eqnarray}
and
\begin{eqnarray}
\mathcal{K}^{i\bar{j}}&=& \frac{1}{\text{det}\,\mathcal{K}_{i\bar{j}}}
\begin{pmatrix}
\mathcal{K}_{\tau\bar{\tau}}&-\mathcal{K}_{\tau\bar{z}}\\
-\mathcal{K}_{z\bar{\tau}}&\mathcal{K}_{z\bar{z}}
\end{pmatrix}\\
\end{eqnarray}

%
%
%

\subsection{Mass matrix near the conifold}
Now we proceed to estimate the values of the mass matrix terms with relation to the distance to the conifold $|z|$ with $z=|z|e^{i\theta}$. We do this while keeping fixed the dilaton value $(\tau-\bar\tau)$  and by considering $c>|z|$. Under these conditions we have that
\begin{equation}
K^0_I=(K^0_i, K^0_{\bar{i}})=
\begin{pmatrix}
K^0_z&K^0_{\bar{z}}\\
K^0_\tau&K^0_{\bar{\tau}}
\end{pmatrix}\approx
\begin{pmatrix}
A\bar{z}\ln|z|^2+B\bar{z}+C& \bar{A}z\ln|z|^2+\bar{B}z\bar{C}\\
-\frac{1}{(\tau-\bar\tau)}&\frac{1}{(\tau-\bar\tau)}
\end{pmatrix}
\end{equation}
and
\begin{equation}
K_I=(K_i, K_{\bar{i}})=
\begin{pmatrix}
K_z&K_{\bar{z}}\\
K_{\tau}&K_{\bar{\tau}}
\end{pmatrix}
=c
\begin{pmatrix}
\frac{e^{i\theta}}{3(\tau-\bar\tau)^2|z|^{1/3}}&\frac{e^{-i\theta}}{3(\tau-\bar\tau)^2|z|^{1/3}}\\
-\frac{2|z|^{2/3}}{(\tau-\bar\tau)^3}&\frac{2|z|^{2/3}}{(\tau-\bar\tau)^3}
\end{pmatrix}
\end{equation}

It is important to notice that in $K_I$ there is an explicit dependence on angle $\theta$ indicating the possibility of obtaining different mass eigenvalues by monodromies.\\

In the following estimations, we make use of
\begin{equation}
\lim_{|z|\rightarrow 0} |z|^m\ln|z|^2=0, \qquad m>0,
\end{equation}
which indicates that near the conifold, the term $1/|z|^m$ dominates over $\ln|z|^2$. With  the purpose to simplify notation, let us parametrize the distance to the conifold by $\epsilon=|z|^{1/3}$. Hence by only keeping
leading  order terms in $\epsilon$, for $\epsilon\ll 1$, we have that\footnote{We do not keep track on signs and constants but only on $|z|$ dependence.}

\begin{equation}
K^0_I\sim
\begin{pmatrix}
\epsilon^3(1+\ln\epsilon^6)+1&\epsilon^3(1+\ln\epsilon^6)+1\\
1&1
\end{pmatrix},
\qquad
K_I\sim c
\begin{pmatrix}
1/\epsilon& 1/\epsilon\\
\epsilon^2&\epsilon^2
\end{pmatrix},
\end{equation}
and
\begin{eqnarray}
K^0_{i\bar{j}}&=&
\begin{pmatrix}
K^0_{z\bar{z}}&K^0_{z\bar\tau}\\
K^0_{\tau\bar{z}}&K^0_{\tau\bar\tau}
\end{pmatrix}=
\begin{pmatrix}
A+B\ln|z|^2 & 0\\
0&\frac{1}{(\tau-\bar{\tau})^2}
\end{pmatrix}
\sim
\begin{pmatrix}
1+\ln\epsilon^6&0\\
0& 1
\end{pmatrix}\\
K^0_{ij}&=&
\begin{pmatrix}
K^0_{zz}&K^0_{z\tau}\\
K^0_{\tau z}&K^0_{\tau\tau}\\
\end{pmatrix}
=
\begin{pmatrix}
Ae^{-2i\theta}+B& 0\\
0&\frac{-1}{(\tau-\bar{\tau})^2}
\end{pmatrix}
\sim
\begin{pmatrix}
1&0\\
0&1
\end{pmatrix}\\
K^0_{\bar{i}\bar{j}}&=&
\begin{pmatrix}
K^0_{\bar{z}\bar{z}}&K^0_{\bar{z}\bar{\tau}}\\
K^0_{\bar{\tau}\bar{z}}&K^0_{\bar{\tau}\bar{\tau}}
\end{pmatrix}
=
\begin{pmatrix}
A+B\ln|z|^2 & 0\\
0&\frac{-1}{(\tau-\bar{\tau})^2}
\end{pmatrix}
\sim
\begin{pmatrix}
1+\ln\epsilon^6&0\\
0& 1
\end{pmatrix},
\end{eqnarray}
where the number 1 indicates a term of order  $\epsilon^0$ but in general a function on $\tau$ and $\bar{\tau}$. Similarly, for the K\"ahler matrices $K_{i\bar{j}}$, $K_{ij}$, $K_{\bar{i}{j}}$ we have that 
\begin{eqnarray}
K_{i\bar{j}}&=&
\begin{pmatrix}
K_{z\bar{z}}&K_{z\bar\tau}\\
K_{\tau\bar{z}}&K_{\tau\bar\tau}
\end{pmatrix}=
\frac{c}{\tau-\bar\tau}
\begin{pmatrix}
\frac{1}{9|z|^{4/3}} & \frac{2e^{-i\theta}}{3|z|^{1/3}(\tau-\bar\tau)}\\
-\frac{2e^{i\theta}}{3|z|^{1/3}(\tau-\bar\tau)}&-\frac{6|z|^{2/3}}{(\tau-\bar\tau)^2}
\end{pmatrix}
\sim
c\begin{pmatrix}
1/\epsilon^4&1/\epsilon\\
1/\epsilon& \epsilon^2
\end{pmatrix}\\
K_{ij}&=&
\begin{pmatrix}
K_{zz}&K_{z\tau}\\
K_{\tau z}&K_{\tau\tau}\\
\end{pmatrix}
=
\frac{-2c}{3(\tau-\bar\tau)^2}
\begin{pmatrix}
\frac{e^{-2i\theta}}{3|z|^{4/3}}& \frac{e^{i\theta}}{(\tau-\bar\tau)^ |z|^{1/3}}\\
\frac{e^{i\theta}}{(\tau-\bar\tau)^ |z|^{1/3}}&\frac{-3|z|^{2/3}}{(\tau-\bar\tau)^2}
\end{pmatrix}
\sim
c\begin{pmatrix}
1/\epsilon^4&1/\epsilon\\
1/\epsilon& \epsilon^2
\end{pmatrix}\\
K_{\bar{i}\bar{j}}&=&
\begin{pmatrix}
K_{\bar{z}\bar{z}}&K_{\bar{z}\bar{\tau}}\\
K_{\bar{\tau}\bar{z}}&K_{\bar{\tau}\bar{\tau}}
\end{pmatrix}
=
\frac{-2c}{3(\tau-\bar\tau)^2}
\begin{pmatrix}
\frac{e^{2i\theta}}{3|z|^{4/3}}& \frac{e^{-i\theta}}{(\tau-\bar\tau)^ |z|^{1/3}}\\
\frac{e^{-i\theta}}{(\tau-\bar\tau)^ |z|^{1/3}}&\frac{-3|z|^{2/3}}{(\tau-\bar\tau)^2}
\end{pmatrix}
\sim
c\begin{pmatrix}
1/\epsilon^4&1/\epsilon\\
1/\epsilon& \epsilon^2
\end{pmatrix}.\\
\end{eqnarray}

The inverse metric $\mathcal{K}^{i\bar{j}}$ is given by
\begin{eqnarray}
\mathcal{K}^{i\bar{j}}&=&
\begin{pmatrix}
\mathcal{K}^{z\bar{z}}&\mathcal{K}^{z\bar\tau}\\
\mathcal{K}^{\tau\bar{z}}&\mathcal{K}^{\tau\bar{\tau}}
\end{pmatrix}=
\frac{1}{\text{det}\,\mathcal{K}_{i\bar{j}}}
\begin{pmatrix}
\mathcal{K}_{\tau\bar\tau}&-\mathcal{K}_{\tau\bar{z}}\\
-\mathcal{K}_{z\bar{\tau}}&\mathcal{K}_{z\bar{z}}
\end{pmatrix}\nonumber\\
&=&
\frac{(\tau-\bar\tau)^3}{K^0_{z\bar{z}}\left((\tau-\bar{\tau})-6c|z|^{2/3}\right)-\frac{2c^2}{9|z|^{2/3}(\tau-\bar{\tau})}+\frac{c}{9|z|^{4/3}}}
\begin{pmatrix}
\frac{(\tau-\bar{\tau})-6c|z|^{2/3}}{(\tau-\bar\tau)^3}&\frac{2c}{3(\tau-\bar\tau)^2}\frac{e^{i\theta}}{|z|^{1/3}}\\
\frac{-2c}{3(\tau-\bar\tau)^2}\frac{e^{-i\theta}}{|z|^{1/3}}&K^0_{z\bar{z}}+\frac{c}{9|z|^{4/3}}
\end{pmatrix},\nonumber\\
&\sim& \frac{9\epsilon^4(\tau-\bar{\tau})^3}{c}
\begin{pmatrix}
\frac{(\tau-\bar{\tau})-6c\epsilon^2}{(\tau-\bar\tau)^3}&\frac{2c}{3(\tau-\bar\tau)^2}\frac{e^{i\theta}}{\epsilon}\\
\frac{-2c}{3(\tau-\bar\tau)^2}\frac{e^{-i\theta}}{\epsilon}&K^0_{z\bar{z}}+\frac{c}{9\epsilon^4}
\end{pmatrix}
\end{eqnarray}
then
\begin{eqnarray}
\mathcal{K}^{i\bar{j}}
&\sim&
\begin{pmatrix}
\epsilon^4&\epsilon^3\\
\epsilon^3&1
\end{pmatrix}\equiv \mathbf{K}.
\end{eqnarray}
In terms of the dependence on $z$ at leading term, we can also write expressions of the matrices $e_{IJ}$. For that, we can consider the superpotential $W$ in terms of $z$ as
\begin{equation}
W\sim a+b z+ c z\ln z +P({\mathcal{O}}(z^p)),
\end{equation}
with $P({\mathcal{O}}(z^p))$ a polynomial on $z$ of order $p$, with $p\geq 2$, and $a,b,c$ constants or functions on $\tau$ and $\bar{\tau}$. Hence, 
\begin{eqnarray}
\partial_z W&\sim& 1+ \ln z + z^{p-1},\nonumber\\
\partial^2 _zW&\sim& \frac{1}{z}+ z^{p-2}.
\end{eqnarray}
Using these estimations we have for instance that
\begin{eqnarray}
e_{zz} &=& \left(\partial_z^2+\mathcal{K}_z\partial_z +\mathcal{K}_{zz}\right)W,\nonumber\\
&\sim&\frac{1}{\epsilon^4}
\end{eqnarray}
such that 
\begin{equation}
e_{ij}\sim e_{\bar{i}j}\sim e_{i\bar{j}}\sim e_{\bar{i}\bar{j}}\sim
\begin{pmatrix}
1/\epsilon^4&1/\epsilon^3\\
1/\epsilon^3&1
\end{pmatrix}\equiv\alpha ,
\end{equation}
With these estimations, near the conifold we can write the matrix mass as
\begin{eqnarray}
(M^I_J)^2&\sim&
\begin{pmatrix}
0&\mathbf{K}\\
\mathbf{K}&0
\end{pmatrix}
\begin{pmatrix}
\alpha &\alpha\\
\alpha&\alpha
\end{pmatrix}
\begin{pmatrix}
0&\mathbf{K}\\
\mathbf{K}&0
\end{pmatrix}
\begin{pmatrix}
\alpha &\alpha\\
\alpha&\alpha
\end{pmatrix} e^{\mathcal{K}}\nonumber\\
&=&2
\begin{pmatrix}
\mathbf{K}\alpha\mathbf{K}\alpha&\mathbf{K}\alpha\mathbf{K}\alpha\\
\mathbf{K}\alpha\mathbf{K}\alpha&\mathbf{K}\alpha\mathbf{K}\alpha
\end{pmatrix}e^{\mathcal{K}}.
\end{eqnarray}

Then, near the conifold, the mass matrix 

\begin{eqnarray}
(M^I_J)^2&=&
\begin{pmatrix}
(M^z_z)^2&(M^z_\tau)^2&(M^z_{\bar{z}})^2&(M^z_{\bar{\tau}})^2\\
(M^\tau_z)^2&(M^\tau_\tau)^2&(M^\tau_{\bar{z}})^2&(M^\tau_{\bar{\tau}})^2\\
(M^{\bar{z}}_z)^2&(M^{\bar{z}}_\tau)^2&(M^{\bar{z}}_{\bar{z}})^2&(M^{\bar{z}}_{\bar{\tau}})^2\\
(M^{\bar{\tau}}_z)^2&(M^{\bar{\tau}}_{\tau})^2&(M^{\bar{\tau}}_{\bar{z}})^2&(M^{\bar{\tau}}_{\bar{\tau}})^2
\end{pmatrix}\nonumber\\
\end{eqnarray}

can be estimated in powers of $\epsilon$ as
\begin{eqnarray}
(M^I_J)^2
&=&
\begin{pmatrix}
0&{\mathcal{K}}^{i\bar{m}}\\
{\mathcal{K}}^{\bar{i}m}&0
\end{pmatrix}
\begin{pmatrix}
V_{mj}&V_{m\bar{j}}\\
V_{\bar{m}j}&V_{\bar{m}\bar{j}}
\end{pmatrix}\\
&\sim&
\begin{pmatrix}
0&0&\epsilon^4&\epsilon^3\\
0&0&\epsilon^3&1\\
\epsilon^4&\epsilon^3&0&0\\
\epsilon^3&1&0&0
\end{pmatrix}
\begin{pmatrix}
1/\epsilon^4&1/\epsilon&1/\epsilon^4&1/\epsilon\\
1/\epsilon&1&1/\epsilon&1\\
1/\epsilon^4&1/\epsilon&1/\epsilon^4&1/\epsilon\\
1/\epsilon&1&1/\epsilon&1
\end{pmatrix} e^{\mathcal{K}}\\
&\sim&
\begin{pmatrix}
1&\epsilon^3&1&\epsilon^3\\
1/\epsilon&1&1/\epsilon&1\\
1&\epsilon^3&1&\epsilon^3\\
1/\epsilon&1&1/\epsilon&1
\end{pmatrix} e^{\mathcal{K}}
\end{eqnarray}

In general we can see some general features.
For example, we  observe that
\begin{equation}
\frac{(M^\tau_\tau)^2}{(M^z_\tau)^2}\sim 1/\epsilon^3,
\end{equation}
meaning that near the conifold $(M^\tau_\tau)^2$ is much larger than $(M^z_\tau)^2$ and can be dismissed. A general set of hierarchies can be inferred as
\begin{eqnarray}
(M^\tau_z)^2\sim (M^{\tau}_{\bar{z}})^2\sim (M^{\bar{\tau}}_z)^2\sim (M^{\bar{\tau}}_{\bar{z}})^2&>& (M^z_z)^2\sim(M^z_{\bar{z}})^2\sim (M^{\bar{\tau}}_\tau)^2\sim(M^\tau_{\bar{\tau}})^2\sim(M^{\bar{\tau}}_{\bar{\tau}})^2\sim (M^\tau_\tau)^2\nonumber\\
&>&(M^z_\tau)^2\sim (M^z_{\bar{\tau}})^2\sim(M^{\bar{z}}_\tau)^2\sim (M^{\bar{z}}_{\bar{\tau}})^2,
\end{eqnarray}
and the mass matrix can be approximated at leading order in $\epsilon$ as
\begin{eqnarray}
(M^I_J)^2&=&
\begin{pmatrix}
(M^z_z)^2&0&(M^z_{\bar{z}})^2&0\\
(M^\tau_z)^2&(M^\tau_\tau)^2&(M^\tau_{\bar{z}})^2&(M^\tau_{\bar{\tau}})^2\\
(M^{\bar{z}}_z)^2&0&(M^{\bar{z}}_{\bar{z}})^2&0\\
(M^{\bar{\tau}}_z)^2&(M^{\bar{\tau}}_{\tau})^2&(M^{\bar{\tau}}_{\bar{z}})^2&(M^{\bar{\tau}}_{\bar{\tau}})^2
\end{pmatrix}.\nonumber\\
\end{eqnarray}
However, notice that this set of hierarchies is clearly a consequence of the correction to the K\"ahler potential given by $K(z,\bar{z},\tau,\bar{\tau})$.  \\

 For
 $K(z,\bar{z},\tau,\bar{\tau})=0$, without  considering a correction in the K\"ahler potential, the matrix $\mathcal{K}^{i\bar{j}}$ is diagonal and
 the mass matrix reads

\begin{equation}
(M^I_J)^2\sim
\begin{pmatrix}
1/\epsilon^3& \ln \epsilon/\epsilon^3&0&\ln\epsilon\\
 \ln \epsilon/\epsilon^3&1/\epsilon^3&\ln\epsilon&1/\epsilon^3\\
0&\ln\epsilon&1/\epsilon^3& \ln \epsilon/\epsilon^3\\
\ln\epsilon&1/\epsilon^3& \ln \epsilon/\epsilon^3&1/\epsilon^3
\end{pmatrix}
\end{equation}
showing a different set of hierarchies among mass terms. For instance, we observe that the K\"ahler correction produces the following change:
\begin{equation}
\frac{(M^\tau_\tau)^2}{(M^z_\tau)^2}\sim\frac{1}{\epsilon^3} \qquad\longrightarrow \qquad\frac{(M^\tau_\tau)^2}{ (M^z_\tau)^2 }\sim\frac{1}{\ln\epsilon},
\end{equation}
inverting the hierarchies among those terms.




\end{appendices}

\bibliographystyle{utphys}
\bibliography{biblioV}

\end{document}